\documentclass[epj]{svjour}
\usepackage{graphics}
\usepackage{graphicx}
\usepackage{amsmath}
\usepackage{color}

\begin{document}
\title{\textit{Ab initio} 
calculations of  optical properties of 
silver clusters: Cross-over from molecular to nanoscale behavior}

\author{
John T. Titantah\inst{1} and Mikko Karttunen\inst{1}
%First author\inst{1} \and Second author\inst{2}% etc
% \thanks is optional - remove next line if not needed
%\thanks{\emph{Present address:} Insert the address here if needed}%
}                     % Do not remove
%
%\offprints{}          % Insert a name or remove this line
%
\institute{
Department of Mathematics and Computer Science \& 
Institute for Complex Molecular Systems,
Eindhoven University of Technology, P.O. Box 513, MetaForum, 5600 MB
Eindhoven, the Netherlands
}
\date{Received: date / Revised version: date}
% The correct dates will be entered by Springer
%
\abstract{
Electronic and optical properties of silver clusters were calculated using two different 
\textit{ab initio} approaches: 
1) based on all-electron full-potential linearized-augmented plane-wave method and 2)  
local basis function pseudopotential approach. Agreement is found between the two methods 
for small and intermediate sized clusters for which the former method is limited due to 
its all-electron formulation. The latter, due to non-periodic boundary conditions, is the
more natural approach to simulate small clusters. The effect of cluster size is then 
explored using the  local basis function approach. We find that as the cluster size increases, 
the electronic structure undergoes a transition from molecular behavior to 
nanoparticle behavior  at a cluster size of 140 atoms (diameter $\sim 1.7$\,nm). 
Above this cluster size the step-like electronic structure, 
evident as several features in the imaginary part  of the polarizability of all clusters 
smaller than Ag$_\mathrm{147}$, gives way to a dominant plasmon peak localized at 
wavelengths 350\,nm$\le\lambda\le$ 600\,nm.  
It is, thus, at this length-scale that the conduction electrons' collective 
oscillations that are responsible
for plasmonic resonances begin to dominate the opto-electronic properties of silver nanoclusters.
 % end of PACS codes
} %end of abstract
\authorrunning{Titantah and Karttunen}
\titlerunning{Ag clusters: crossover to nanoscale behavior}

\maketitle

\section{Introduction}

The readiness with which noble metal nanoparticles create and support surface plasmons makes them
useful in a wide range of applications, e.g., in photonic devices, as chemical 
sensors~\cite{chen09,filippo09,he10,nezhad10,ngece11}, in bio-imaging~\cite{kravets12}, 
in drug delivery~\cite{portney06}, cancer therapy~\cite{portney06,huang06,shenoi13}, optical 
manipulation~\cite{ashkin70,ashkin86,liu10,filippo09}, electrical conduits in microelectronic 
industry~\cite{chendapeng09,hsu07,lee05b}, and even as 'nanoears' in
optical readout of acoustic waves generated in liquid media~\cite{olinger12}
and in channeling photon energy in vortex nanogear transmission~\cite{boriskina12}. 

The changing electronic and optical properties of these materials with size from molecular clusters 
through nanometer  scale to micrometer range and their morphology govern their domain of applicability. 
However, the properties of the sub-10 nm end of these noble metals, for which quantum effects become 
crucial~\cite{scholl12}, is not well studied although this end is of great importance for biological 
applications. Beyond this range  theoretical studies of the optical properties are usually conducted 
based on classical electrodynamic approaches. Medium sized clusters (20-500 atoms) have scarcely been 
investigated, whereas at this size scale intriguing behaviors have been reported. For instance, 
Yuan and coworkers observed an unexpected luminescence change during the purification of 
thiolate-protected $<$2 nm Ag nanoclusters; the non-luminescent Ag nanocluster became highly luminescent 
when passed through a separation column~\cite{yuan13}. Full quantum investigations are often limited 
to few atoms clusters. These involve tight-binding calculations~\cite{bassani85,bifone94,burgess11}, 
and the time-dependent density functional theory (TDDFT) approach~\cite{burgess11}.  A promising atomistic 
method that can be conveniently used on  hundreds of  atom systems is the capacitance-polarizability 
interaction model~\cite{jensen01,jensen09} which has been successfully used to study the 
effect of planar defects on the optical properties of Ag nanostructures~\cite{ben2013}. Those based 
on the classical electrodynamic theory of a dielectric object in an electromagnetic field use methods 
such as the Mie theory~\cite{mie1908,miljkovic10}, the discrete dipole approximation~\cite{draine88,draine94} 
and the finite-difference time-domain approaches. Optical forces are also often computed using 
the Maxwell stress tensor~\cite{okamoto99,miljkovic10}.  All these classical electrodynamic studies 
exploit the frequency-dependent dielectric constant of the bulk materials, the size dependence of  
these dielectric constants being extrapolated through the size dependence of the life-time broadening 
$\gamma=Av_F/R$ which requires the knowledge of the Fermi velocity $v_F$ of the bulk metal.  The lower 
limit of the dimensions of the nanoparticle that this extrapolation may apply to is dictated by the 
electron mean free path in the bulk system of $\sim$57 nm~\cite{xu05}. A thorough DFT study on spherical  
Ag clusters of sizes ranging from 0.4 to  2\,nm have found that the plasmon frequency becomes 
size-dependent and that the collision frequency $\gamma$ has a more complex dependence on the particle 
radius~\cite{he10} than the usual 1/$R$ dependence.  Because of this limitation, the usual size-dependent 
Drude-model may not be applicable to small nanoparticles where quantum confinement of the electrons become 
very important~\cite{scholl12}, and certainly may not help in understanding the 
mechanism governing the optical properties of nanoparticles and clusters.

In this work, we use two quantum mechanical approaches to calculate the complex permittivity of 
Ag clusters with sizes ranging from few atoms to hundreds of atoms. The two methods used are 1) the 
all-electron full-potential linearized-augmented plane-wave~\cite{wien2k} and 2)  
{ local basis function DFT approach} ~\cite{siesta97,siesta02}.  
The latter has been used to predict the half-metallic character of 
graphene nanoribbons~\cite{young-woo06}  
{and to probe the optical properties of armchair 
graphene nanoribbons embedded in hexagonal boron nitride lattices~\cite{nematian12}}. 
{ The local basis function approach}  is adopted for further studies due to its simplicity and  robustness.  
 
\section{Methods}

We used the WIEN2K all-electron-full-potential linearized-augmented plane-wave (LAPW) code~\cite{wien2k},  
with the generalized gradient approximation (GGA)~\cite{gga96}  for the exchange and correlation potential,
to calculate the electronic and optical properties of Ag clusters. In this method, the system 
is described by a 3D periodically repeated unit cell which is partitioned into non-overlapping 
muffin-tin spheres centered on each atom and interstitials. The Kohn-Sham wave function is described as 
a linear combination of atomic basis functions in the muffin-tin spheres and as plane-waves in the interstitials. 
{Muffin-tin radii of 1.3~\AA~around the Ag atoms are used. The multipolar Fourier expansion of  
Weinert \textit{et al.}~\cite{weinert82} is used to compute the Coulomb potential.}
{Using the LAPW method, the cluster of interest is positioned at the center of a cubic unit cell with sides 
larger than the diameter of the cluster}. 
The limit of isolated cluster cannot be reached using this approach 
{since (strictly speaking) infinitely large unit cell is required and} the computation in empty 
space becomes prohibitive for large unit cells; this method can only give trends towards isolated clusters.
Two parameters govern the accuracy of the calculations: 1) The product of the plane-wave cut-off and the smallest 
muffin-tin  radius in the system ($RKM$), and 2) the  number of $k$-points in the irreducible Brillouin zone (IBZ).  
 For these clusters, $RKM$-value of 5.5 is sufficient to yield converged optical properties. 
We use 4 $k$-points in the IBZ for the self-consistent calculations, and 20 for calculations of optical properties.

We also employed a second method,   
a { local basis function approach} within the generalized 
gradient approximation of the exchange and correlation energy 
{of Perdew, Burke and  Wang~\cite{gga96}} as implemented in 
the SIESTA code~\cite{siesta97,siesta02}.  
{ In this method the  basis functions are made to vanish beyond a
predefined cut-off distance.}
{For clusters, a multi-grid approach was used to calculate the 
electrostatic potential in accordance with the range of the lower moments of the charge density.}
{This approach is particularly appropriate for large clusters 
as it scales linearly with the number of atoms in the cluster.}
The core-electrons are described by the Troullier-Martins norm conserving pseudopotential~\cite{troullier91}. 
The 4d$^{10}$5s$^1$ electrons are considered as the valence electrons for Ag. The double-$\zeta$  
(DZ) basis set~\cite{davidson86} (each occupied orbital is described by two basis functions) 
is used to represent the valence states.  
{ No polarization function was used as attempts to include it did not improve the results.}
{With LAPW, periodic boundary conditions were applied. The SIESTA calculations 
did not use them.}

{For the  LAPW approach optical properties were obtained from the joint density of states modified
with the respective dipole matrix elements~\cite{ambrosch95}. The SIESTA code computes 
the imaginary part of the dielectric function using the linear response function~\cite{yu01}:}
 \begin{eqnarray}
\epsilon_\mathrm{2}(\omega)&=&{1\over 4\pi\epsilon_\mathrm{0}}
\left({2\pi e\over m\omega}\right)^2\sum_{{\bf k}}\mid{\bf p_{c,v}}\mid^2\delta
\left(E_\mathrm{c}({\bf k})-E_\mathrm{v}({\bf k})-\hbar\omega\right)\nonumber \\
&&\times\left[f\left(E_\mathrm{v}({\bf k})\right)-f\left(E_\mathrm{c}({\bf k})\right)\right],
\end{eqnarray}
{where the subscripts $v$ and  $c$ represent the valence and conduction bands, respectively, 
$\hbar\omega$ the photon energy,  $m$ the electron mass, $E_\mathrm{c,v}$({\bf k}) the conduction 
and valence band energies with $k$-vector ${\bf k}$, and {$\bf p_\mathrm{c,v}$} is the momentum operator. 
$f$ is the Fermi function and $\delta$ is the Dirac delta function.
The real part of the dielectric function was obtained by using the 
Kramers-Kronig relation }
\begin{equation}
 \epsilon_\mathrm{1}(\omega)=1+{2\over \pi}P\int_0^\infty
{\omega^\prime\epsilon_\mathrm{2}(\omega^\prime)\over {\omega^\prime}^2-\omega^2}d\omega^\prime,
\end{equation}
where $P$ stands for the Cauchy principal value.

Relaxed clusters were used for studying optical properties. Relaxation was performed as follows: Each cluster 
was cut out from an FCC structure  and 
pre-relaxed using a molecular dynamics (MD) run by using a recently parameterized many-body Gupta potential 
for Ag~\cite{titantah-param}.  The MD relaxation was performed at 300\,K with the temperature kept constant 
using the Nos\'e-Hoover thermostat~\cite{nose-molphys84,hoover85}. The resulting structures were further 
relaxed using the SIESTA code~\cite{siesta97,siesta02} via the conjugate gradient method.
The DFT fine-tuning was continued until forces on all atoms became smaller than 40 meV/{\AA}. 

\section{Results} 

\subsection{Comparison between   local basis function  and plane wave methods}

He and Zeng~\cite{heyi10} used the SIESTA code~\cite{siesta97,siesta02} to 
calculate the optical properties of Ag clusters 
{with sizes ranging from 13 to 586 atoms}
and noted the need to scissor-shift their spectra by a fixed energy of 1.28\,eV to align with experimental results; 
{when comparing different computational results, it is important to notice that the magnitude of the 
shift needed to align with experiments
depends on the exchange correlation functional. }
The need to apply a scissor-shift is a consequence of the well-known underestimation of the band-gap by the single 
particle DFT approach~\cite{perdew85,HybertsenLouie1986,heyi10};
techniques that contain the missing many-body effect like the GW (BSE)
approach~\cite{hedin65,HybertsenLouie1986}  cannot be used with large system sizes such as here, 
since the computational cost scales as $N^4$, where $N$ is the number of electrons in the system. 

\begin{figure}
%\begin{center}
\includegraphics[width=8cm]{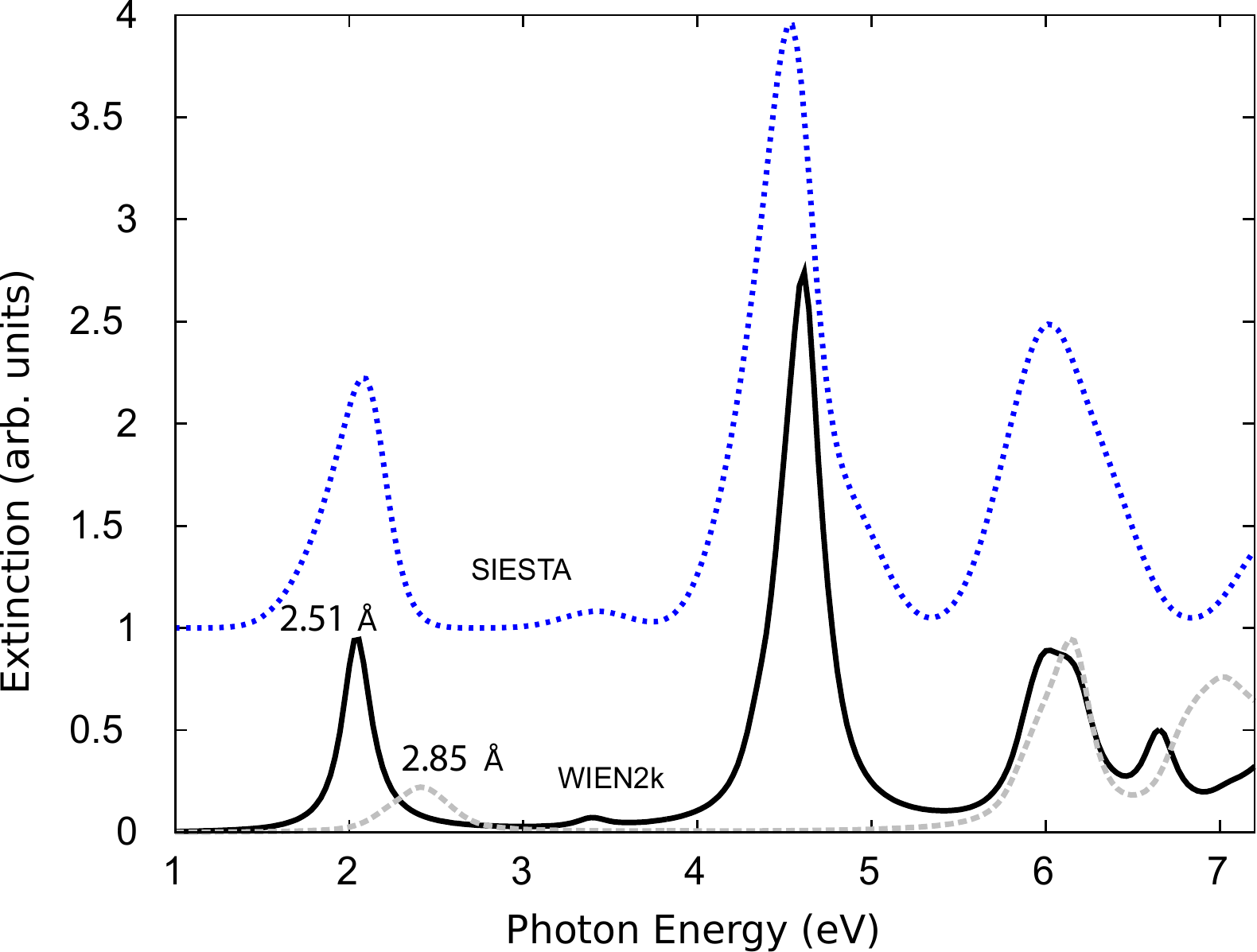}
  \caption{Comparison of the extinction coefficient of Ag$_2$ calculated using the SIESTA and the WIEN2k codes. 
  All optical peaks are well reproduced by both methods apart from a global redshift of about 0.3\,eV  
  of the WIEN2k results with respect to the SIESTA one. Thus the WIEN2k spectrum has been shifted by 
  0.3\,eV to align with  
 local basis function  calculations. 
 For WIEN2k, we show results for two dimer bond lengths of 2.51~{\AA} (black) and 2.85~{\AA} (light grey) 
 while the SIESTA result is given only for the equilibrium dimer bond length of 2.51~{\AA}.
}
\label{fig:dimer}
%\end{center}
\end{figure}

{Since one of the aims here is to compare the applicabilities of the two different
approaches, the results presented in this work do not include any shift except for  $\sim$0.3 eV blue-shift to 
the WIEN2k data to align with the SIESTA results for easier comparison}. 
Our calculations of the extinction coefficient of the Ag dimer (Ag$_2$) revealed four major features 
within the energy range of 1-7\,eV, Fig.~\ref{fig:dimer}.  
All optical peaks are well reproduced by both methods apart from a 
global redshift of about 0.3\,eV  of the WIEN2k results with respect to the SIESTA ones. This shift is 
most likely a consequence of the  periodic boundary conditions in  the WIEN2K simulations (a cubic unit 
cell with side length of 10~{\AA}).  
{The equilibrium dimer bond length of 2.51~{\AA} was found, which compares very well 
with the measured value of 2.53~{\AA}\,\cite{simard91} and another computed value 
of 2.52~{\AA}~\cite{martin87}.
}

Two of the features (at 1.8-2.4\,eV and 4.3\,eV) 
are  sensitive to the Ag-Ag bond length:  A computation using a  relatively long dimer bond length 
of 2.85~{\AA} is seen to affect these features, Fig.~\ref{fig:dimer}.
These features may be responsible for the widely varying 
optical spectra~\cite{Idrobo07} of Ag nanoclusters consisting of only a few atoms since such small 
clusters may have a wide 
variation in bond lengths. The optical features of Ag$_2$ dimer found  here are consistent with 
the density of states (DOS) calculation of Pereiro \textit{et al.}~\cite{pereiro07};
the imaginary part of the dielectric constant is proportional to the valence- and 
conduction-band joint density of states. 
The minor feature at 3.5\,eV is again reproduced by both techniques.

The robustness of SIESTA approach is further demonstrated by the optical absorption 
of triangular Ag$_3$ that yields four features at 320-350,   360-390, 430-460  
and 460-500\,nm. These numbers compare well with the experimental values of 320-330, 380-390, 
410-420 and 480-490\,nm as reported in Ref.~\cite{fedrigo93}.
Next, we compare the absorption coefficients for medium sized  clusters.

\begin{figure}
%\begin{center}
\includegraphics[width=8cm]{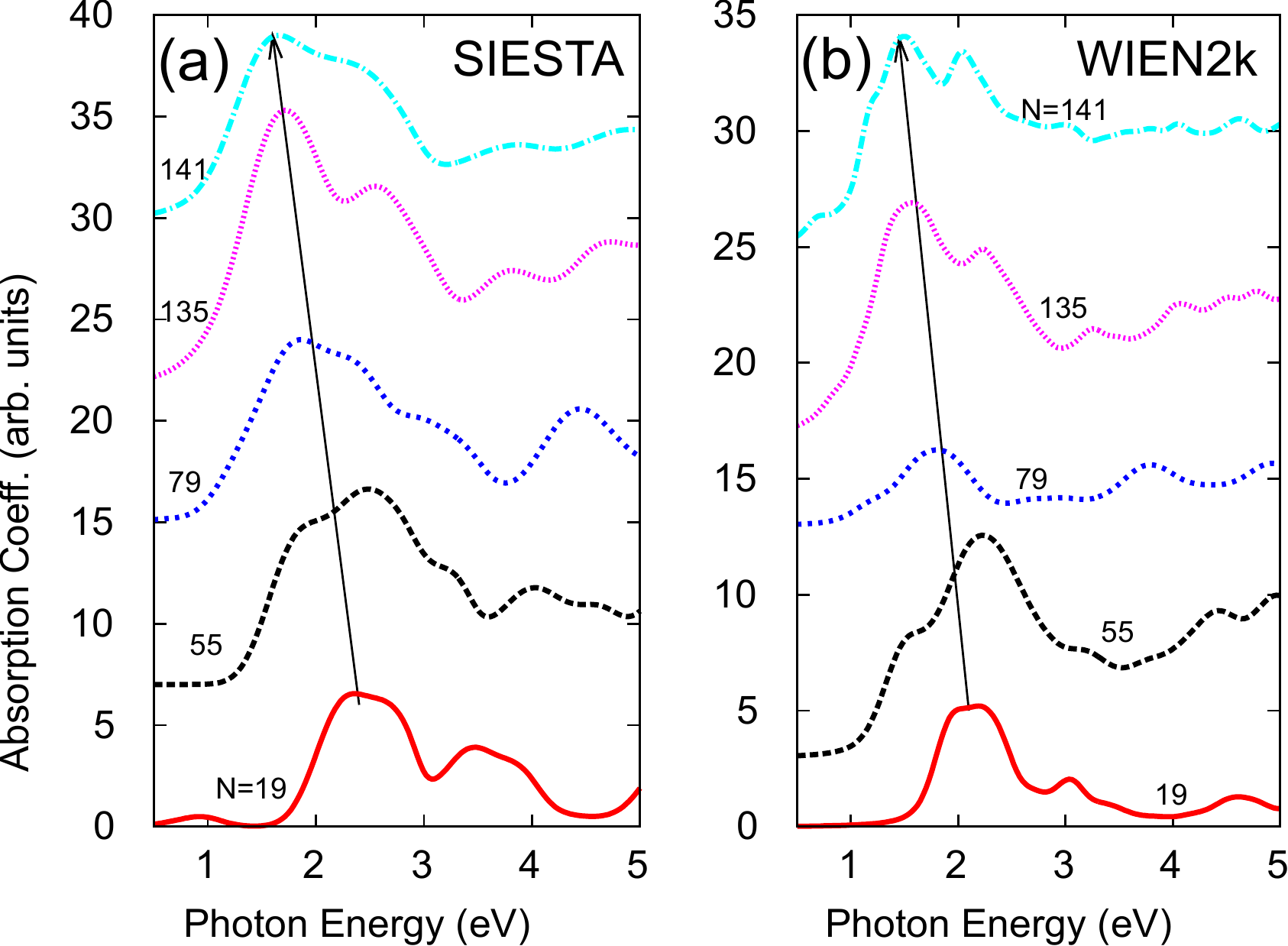}
  \caption{Absorption coefficients. %of \protect{Ag$_N$} clusters calculated using a) the SIESTA and b) the WIEN2k  codes. 
Lines: Tendency of the absorption onset to shift to lower 
  energies as cluster size increases.}
\label{fig:absorp-n}
%\end{center}
\end{figure}

Figure~\ref{fig:absorp-n} shows the absorption coefficient on a series of Ag clusters 
with size ranging from 19 to 141 atoms. The methods compare well with each other in terms 
of the peak positions (aside from a global redshift of 0.3 eV of the WIEN2k spectra with respect to 
the SIESTA's) and the relative heights of the peaks. 
{ The results for Ag$_\mathrm{55}$  compare very well with 
measured UV-photoelectron spectra on Ag$_\mathrm{55}^{-}$~\cite{hakkinen04}, 
apart  from the global shift to lower photon energy of 1.3~eV 
(practically the value of the scissor-shift of 1.28 eV reported by 
He and Zeng~\cite{he10}). Both the position and relative intensities of the 
four features within 5~eV agree well with UV-photoelectron data. 
The comparison appears even better than with very recent 
TDDFT calculations~\cite{kuisma15}.}

As the next step, we had an Ag$_{55}$ dimer at varying separations ranging from touching to 
about 10\,nm. Figure~\ref{fig:dimer-ext}(a) reveals a transition from a sharp to  
a broad resonance. This is very similar to the experimentally 
reported transition 
when an array of 100\,nm Ag nanoclusters embedded 
into poly (dimethylsiloxane) (PDMS) was stretched to increase the inter-particle separation~\cite{evanoff05}. 
This transition was attributed to quadrupolar interactions which dominate for such large nanoparticles. 
Quadrupolar interactions are, however,
negligible for the cluster sizes considered here. This indicates that dipolar interactions also 
contribute to this effect. 

{In the LAPW approach which uses periodic boundary conditions within a cubic unit cell, the size 
of the unit cell determines the inter-cluster distance. 
On increasing the unit cell size for the Ag$_{55}$ calculation (equivalent to increasing 
inter-cluster separation), 
we find that the same trend of transition from a sharp to a broad resonance, 
as found using SIESTA (see Fig.~\ref{fig:dimer-ext}A),
is also present in the WIEN2k results. For example, as the unit cell size increases from 1.2 to 2.5\,nm, the feature at 370\,nm 
diminishes and eventually vanishes leaving a single peak which shifts from 480 to 520\,nm as seen in Fig.~\ref{fig:dimer-ext}B.

The redshift in the plasmon resonance frequency with increasing particle separation is also in 
agreement with experiments~\cite{kinnan10} and theoretical calculations based on the 
discrete dipole approximation on dimers~\cite{lin11}. 
Our results also confirm that the plasmon resonance becomes narrower with increasing  cluster 
size as will be shown later~\cite{sendova06}. We also studied the effect of unit cell size 
on the extinction coefficient of Ag$_{19}$ ($2R\sim$ 1\,nm). For large unit cells ($2R/L< 1/2$),  
using WIEN2k, the spectrum is characterized by a single dominant feature at about 530\,nm. But for small unit cells
a second peak appears at a shorter wavelength which may indicate the appearance 
of conductive overlap~\cite{atay04} between even such small clusters.   

The above results show that both the SIESTA and WIEN2k approaches yield qualitatively similar 
optical properties. Considering the good agreement between the two methods, we use  the
{ local basis 
function}  method in further studies below.

\begin{figure}
%\begin{center}
\includegraphics[width=8cm]{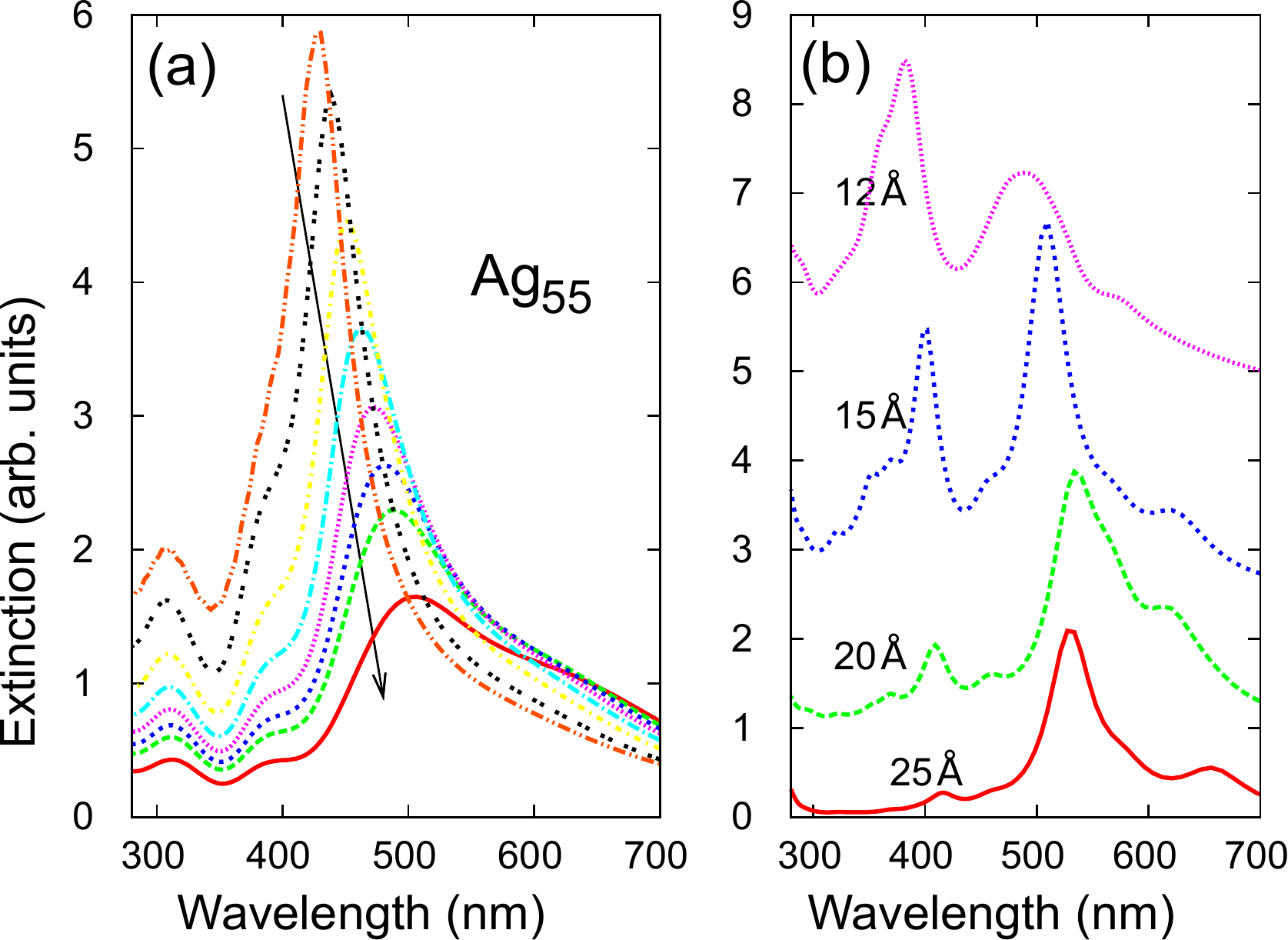}
  \caption{Extinction coefficient of Ag$_{55}$ dimers  as obtained 
  from a) SIESTA code and b) a 3D array of Ag$_{55}$ clusters using WIEN2k 
  for various unit cell sizes  ranging from 12~{\AA}~ to 25~{\AA}.  The arrow points 
  towards increasing dimer separation from touching (center-to-center distance of 13~{\AA}) 
  to well separated clusters ($>$10 nm).  In b) notice that as the unit cell size increases, the intensity 
  of the peak at 370-400\,nm vanishes while that at 500\,nm increases.
}
\label{fig:dimer-ext}
%\end{center}
\end{figure}

\subsection{Plasmon energy and cluster size} 

{Using the SIESTA code,} 
complex polarizability $\alpha$ was calculated for cluster sizes ranging from 55 to 369 atoms,
Fig.~\ref{fig:polariz-N}. 
The Clausius-Mossotti  
polarizabilities 
were calculated using
\begin{equation}
\alpha(\omega)={3d^3 \over 4\pi}{\epsilon(\omega)-\epsilon_0 \over \epsilon(\omega)+2\epsilon_0 },
\label{eq:clausius}\end{equation}
{where $d$ is the size of the cluster,  $\epsilon_0$ the dielectric constant 
of the surrounding medium and $\epsilon(\omega)$ is the frequency dependent complex dielectric constant of the cluster.} 
The plasmon resonances show  a crossover which  appears 
at the cluster size of about 140. 
{For N$<$140 the molecular character shows up as 
multiple features in the imaginary part of the polarizability of smaller clusters.  
For example, the two features at $\sim$3\,eV and 3.5\,eV 
in the imaginary part of the polarizability of Ag$_\mathrm{87}$ and Ag$_\mathrm{135}$ merge into one 
at 3-3.5\,eV for clusters above 140 atoms. For the smaller Ag$_\mathrm{55}$ cluster, the skewed shape of the imaginary part of the 
polarizability is a sign of more than one feature in the optical spectrum. For clusters larger than Ag$_\mathrm{135}$ only one 
symmetric feature, the plasmon peak, can be seen.} 

The cluster size at which this crossover is found is similar to that reported for $<$2\,nm 
Ag nanoclusters~\cite{yuan13}. 
For $N>140$, the molecular transitions give way to nanometallic single optical transition energy which is 
characteristic of s-electrons excitations of the noble metal.  This finding is in 
excellent agreement with TDDFT computations of Weissker \textit{et al.} who found a 
cross-over from multi-feature optical spectrum of Ag clusters as the size increases 
above 140 atoms~\cite{weissker11}. 
{It is also in good agreement with a recent UV/VIS  study on silver clusters
protected with 2-phenylethanethiol (PET), 4-flurothiophenol (4-FTP) and (4-(t-butyl)benzenethiol (BBS) 
which found the emergence of metallicity in Ag clusters composed of 150 atoms~\cite{chakraborty14}}. 
We also find that the optical transitions  redshift as the cluster size increases.  
The tendency of red-shifting with increasing cluster size is also evident in the WIEN2K 
results, 
Fig.~\ref{fig:absorp-n}b. This redshift has also 
recently been reported for glutathione-stabilized magic numbers 
{(15, 18, 22, 25, 29 and  38)} 
1-2.5\,nm Ag clusters~\cite{kumar10} 
whereby a peak-edge onset of 2.1\,eV for Ag$_{15}$  was found. This agrees with the value of 
2.3\,eV  (530\,nm) that we found for Ag$_{19}$. Recent studies using surface second 
harmonic generation spectroscopy~\cite{lunskens15} and manufactured highly fluorescent 
Ag nanoclusters in alumina-silica composite  optical fiber~\cite{halder15} also 
found redshift with increasing cluster size. Similar redshift has also been found for Au clusters~\cite{zheng07}. 
{The observation of redshift with increasing cluster size for  small 
nanoclusters (where higher multipoles than dipole effects are absent) as the ones studied here, 
indicates that the increasing importance of higher multipole effects with increasing  nanoparticle size 
is not enough to explain alone the redshift as has been previously suggested~\cite{link00}}.

\begin{figure}
%\begin{center}
\includegraphics[width=8cm]{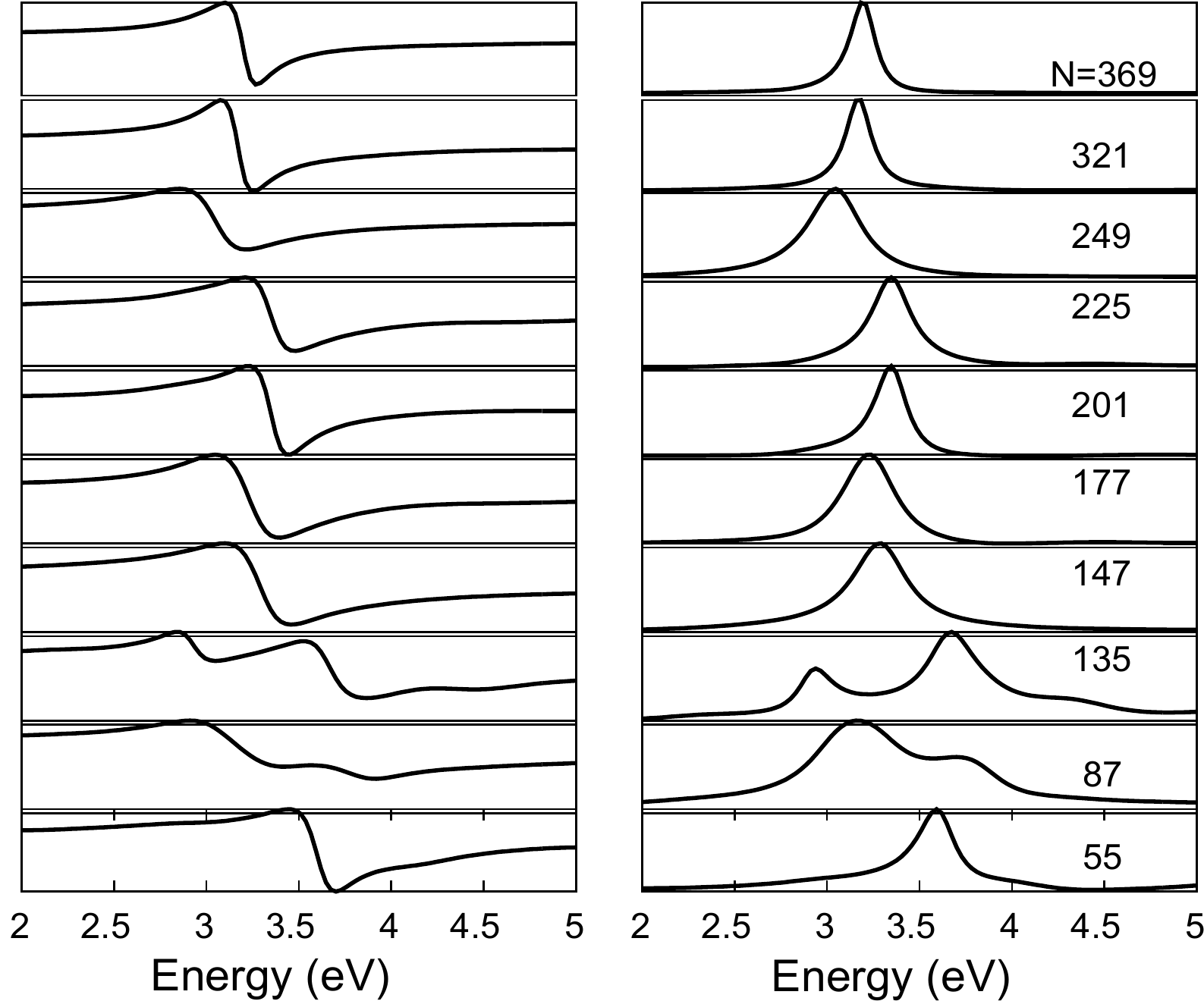}
  \caption{Cluster size effect on the complex polarizability of Ag clusters 
  for cluster sizes ranging from 55 to 369 atoms. The real part is shown in 
  the left and the imaginary in the right. 
{Note that for clusters smaller than Ag$_\mathrm{147}$, 
  the imaginary part has more than one optical feature - a signature of molecular character.}}
\label{fig:polariz-N}
%\end{center}
\end{figure}

\subsection{Effect of cluster shape on the optical properties}
In order to understand the effect of shape change on the optical behavior of Ag clusters, 
we elongated an almost spherical Ag$_{369}$ cluster into an almost elliptical one.
The resulting structures were relaxed using 
DFT  and the optical properties were evaluated. Figure~\ref{fig:opt-ellips} 
shows the wavelength dependence of the optical absorption of these deformed clusters. 
The approximate major to minor axis ratio is denoted by $R^\mathrm{maj/min}$. 
For large nanoparticles, we would expect the optical properties 
to become anisotropic, with each optical spectrum revealing two absorption modes as the 
cluster adopts cigar or rod-like shapes: Longitudinal (low energy) and transverse (high energy) 
modes.  The calculated optical absorption of Ag$_{369}$, for {$R=1^\mathrm{maj/min}$}  clearly shows 
two features: One at 500\,nm 
and the other at about 800\,nm, revealing a not-so-spherical shape. 
The latter wavelength falls 
well within reported values 
{of 700-1100 nm} for the longitudinal plasmon resonance of Ag nanorice
(rounded nanobars)~\cite{wiley07}. 
Upon deforming further this structure 
to get an approximate aspect ratio of $R^\mathrm{maj/min}=$2, the intensity of the 500\,nm feature reduces 
and broadens, and the wavelength  of the peak at 800\,nm only slightly increases. At an 
aspect ratio of about 3 a major shift in the longitudinal mode to a longer wave length  
of $\sim$1100 nm is obtained. This clearly indicates that even at such particle sizes, 
shape plays a crucial role for localized surface plasmons of Ag clusters. However, 
of the three structures of  Ag$_{369}$ studied here,  the ground state energy of the 
undeformed is the lowest, clearly showing that small clusters prefer spherical shapes.

\begin{figure}
%\begin{center}
\includegraphics[width=8cm]{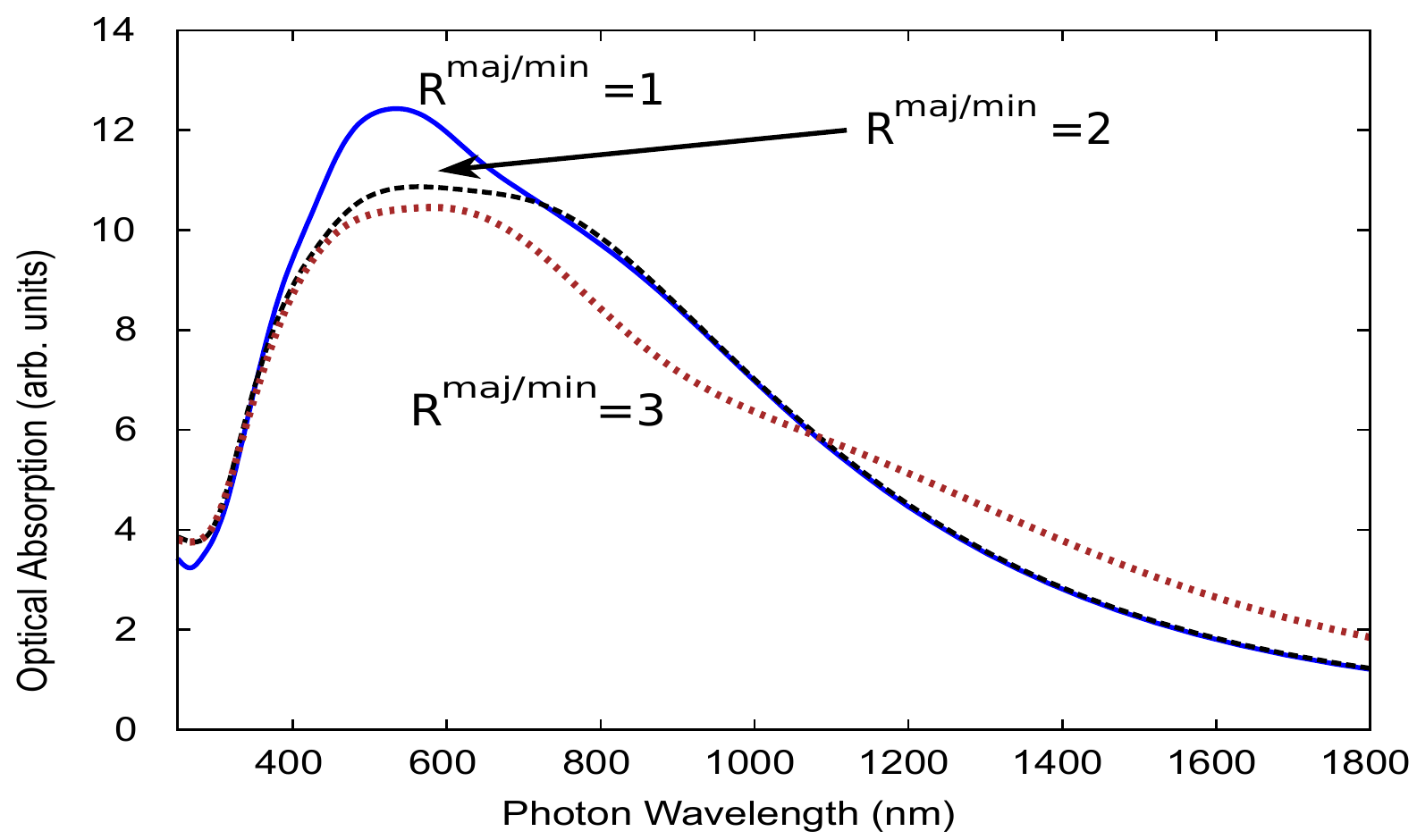}
  \caption{The optical absorption of spherical and elliptical  Ag  clusters. 
  The aspect ratio $R^\mathrm{maj/min}$ of the clusters of similar number of atoms Ag$_{369}$ are indicated 
  as approximate ratio of the principal to the minor axis. }
\label{fig:opt-ellips}
%\end{center}
\end{figure}

\section{Summary}

Two different DFT approaches were used to compute the optical and electronic  
properties of small and medium sized Ag clusters with reasonable agreement between them.  
Not suprisingly, the local basis function approach using SIESTA code turned out to be more appropriate for 
optical properties of Ag nanoclusters and it provides a very good description of 
their structural properties. 
Using this method, we locate a cross-over to nanoscale behaviour at the size of 140 atoms. At this size, the electronic 
properties of Ag clusters change from  molecular character (discrete transition energies) 
to a nanoparticle behavior characterized by a single plasmon resonance frequency. We 
demonstrated the tendency of redshifting when clusters begin to agglomerate to form composite 
structures. 

The results of  the local basis function  DFT computations 
demonstrate the power of the approach and its potential in understanding the structural, 
energetic, electronic and optical properties of technologically relevant noble metal nanoparticles 
and their derivatives. 

\section*{Acknowledgments}
We thank Bj\"orn Baumeier for critical reading of the manuscript.
The initial part of this work was supported by the Natural Sciences 
and Engineering Research Council of Canada (MK).

%
% BibTeX users please use
%  \bibliographystyle{epj}
% \bibliography{}

\begin{thebibliography}{69}

\bibitem{chen09}
Y.Q. Chen, C.J. Lu, Sens. Actuators B Chem. \textbf{135}, 492 (2009)

\bibitem{filippo09}
E.~Filippo, A.~Serra, D.~Manno, Sens. Actuators B Chem. \textbf{138}, 625
  (2009)

\bibitem{he10}
X.~He, C.~Hu, H.~Liu, G.~Du, Y.~Xi, Y.~Jiang, Sens. Actuators B Chem.
  \textbf{144}, 289 (2010)

\bibitem{nezhad10}
M.R.H. Nezhad, J.~Tashkhourian, J.~Khodaveisi, J. Iran. Chem. Soc. \textbf{7},
  S83 (2010)

\bibitem{ngece11}
R.F. Ngece, N.~West, P.M. Ndangili, R.A. Olowu, A.~Williams, N.~Hendricks,
  S.~Mailu, P.~Baker, E.~Iwuoha, Int. J. Electrochem. Sci. \textbf{6}, 1820
  (2011)

\bibitem{kravets12}
V.V. Kravets, K.~Culhane, I.M. Dmitruk, A.O. Pinchuk, \emph{Glycine-coated
  photoluminescent silver nanoclusters}, in \emph{Colloidal Nanocrystals for
  Biomedical Applications VII}, edited by W.~Parak, K.~Yamamoto, M.~Osinski
  (SPIE, 2012), Vol. 8232 of \emph{Proceedings of SPIE}, ISBN 978-0-8194-8875-6

\bibitem{portney06}
N.~Portney, M.~Ozkan, Anal. Bioanal. Chem. \textbf{384}, 620 (2006)

\bibitem{huang06}
X.~Huang, I.~El-Sayed, W.~Qian, M.~El-Sayed, J. Am. Chem. Soc. \textbf{128},
  2115 (2006)

\bibitem{shenoi13}
M.M. Shenoi, I.~Iltis, J.~Choi, N.A. Koonce, G.J. Metzger, R.J. Griffin, J.C.
  Bischof, Molec. Pharmaceutics \textbf{10}, 1683 (2013)

\bibitem{ashkin70}
A.~Ashkin, Phys. Rev. Lett. \textbf{24}, 156 (1970)

\bibitem{ashkin86}
A.~Ashkin, J.M. Dziedzic, J.E. Bjorkholm, S.~Chu, Optics. Lett. \textbf{11},
  288 (1986)

\bibitem{liu10}
M.~Liu, T.~Zentgraf, Y.~Liu, G.~Bartal, X.~Zhang, Nature Nanotechnol.
  \textbf{5}, 570 (2010)

\bibitem{chendapeng09}
D.~Chen, X.~Qiao, X.~Qiu, J.~Chen, J. Mater. Sci. \textbf{44}, 1076 (2009)

\bibitem{hsu07}
S.L.C. Hsu, R.T. Wu, Mater. Lett. \textbf{61}, 3719 (2007)

\bibitem{lee05b}
H.~Lee, K.~Chou, Z.~Shih, Int. J. Adhes. Adhes. \textbf{25}, 437 (2005)

\bibitem{olinger12}
A.~Ohlinger, A.~Deak, A.A. Lutich, J.~Feldmann, Phys. Rev. Lett. \textbf{108}
  (2012)

\bibitem{boriskina12}
S.V. Boriskina, B.M. Reinhard, Nanoscale \textbf{4}, 76 (2012)

\bibitem{scholl12}
J.A. Scholl, A.L. Koh, J.A. Dionne, Nature \textbf{483}, 421 (2012)

\bibitem{yuan13}
X.~Yuan, Q.~Yao, Y.~Yu, Z.~Luo, X.~Dou, J.~Xie, J. Phys. Chem. Lett.
  \textbf{4}, 1811 (2013)

\bibitem{bassani85}
F.~Bassani, M.~Bourg, F.~Cocchini, II Nuovo Cimento D \textbf{5}, 419 (1985)

\bibitem{bifone94}
A.~Bfone, F.~Bassani, Z. Phys. D \textbf{29}, 73 (1994)

\bibitem{burgess11}
R.W. Burgess, V.J. Keast, J. Phys. Chem. C \textbf{115}, 21016 (2011)

\bibitem{jensen01}
J.~Kongsted, A.~Osted, L.~Jensen, P.~Astrand, K.~Mikkelsen, J. Phys. Chem. B
  \textbf{105}, 10243 (2001)

\bibitem{jensen09}
L.L. Jensen, L.~Jensen, J. Phys. Chem. C \textbf{113}, 15182 (2009)

\bibitem{ben2013}
X.~Ben, P.~Cao, H.S. Park, J. Phys. Chem C \textbf{117}, 13738 (2013)

\bibitem{mie1908}
G.~Mie, Ann. Phys. (Leipzig) \textbf{25}, 377 (1908)

\bibitem{miljkovic10}
V.D. Miljkovic, T.~Pakizeh, B.~Sepulveda, P.~Johansson, M.~Kall, J. Phys. Chem.
  C \textbf{114}, 7472 (2010)

\bibitem{draine88}
B.T. Draine, Astrophys. J. \textbf{333}, 848 (1988)

\bibitem{draine94}
B.T. Draine, P.J. Flatau, J. Opt. Soc. Am. A Opt. Image Sci. Vis. \textbf{11},
  1491 (1994)

\bibitem{okamoto99}
K.~Okamoto, S.~Kawata, Phys. Rev. Lett. \textbf{83}, 4534 (1999)

\bibitem{xu05}
H.~Xu, Appl. Phys. Lett. \textbf{87}, 066101 (2005)

\bibitem{wien2k}
P.~Blaha, K.~Schwarz, P.~Sorantin, S.B. Trickey, Comput. Phys. Commun.
  \textbf{59}, 399 (1990)

\bibitem{siesta97}
D.~Sanchez-Portal, P.~Ordejon, E.~Artacho, J.~Soler, Int J Quantum Chem
  \textbf{65}, 453 (1997)

\bibitem{siesta02}
J.~Soler, E.~Artacho, J.~Gale, A.~Garcia, J.~Junquera, P.~Ordejon,
  D.~Sanchez-Portal, J. Phys. Condens Matter \textbf{14}, 2745 (2002)

\bibitem{young-woo06}
Y.W. Son, M.L. Cohen, S.G. Louie, Nature \textbf{444}, 347 (2006)

\bibitem{nematian12}
H.~Nematian, M.~Moradinasab, M.~Pourfath, M.~Fathipour, H.~Kosina, J. Appl.
  Phys. \textbf{111}, 093512 (2012)

\bibitem{gga96}
J.~Perdew, K.~Burke, Y.~Wang, Phys. Rev. B \textbf{54}, 16533 (1996)

\bibitem{weinert82}
M.~Weinert, E.~Wimmer, A.J. Freeman, Phys. Rev. B \textbf{26}, 4571 (1982)

\bibitem{troullier91}
N.~Troullier, J.L. Martins, Phys. Rev. B \textbf{43}, 1993 (1991)

\bibitem{davidson86}
E.R. Davidson, D.~Feller, Chem. Rev. \textbf{86}, 681 (1986)

\bibitem{ambrosch95}
C.~Ambrosch-Draxl, J.A. Majewski, P.~Vogl, G.~Leising, Phys. Rev. B
  \textbf{51}, 9668 (1995)

\bibitem{yu01}
P.T. Yu, M.~Cardiona, \emph{Fundamentals of Semiconductors: Physics and
  Materials Properties} (Springer, Berlin, 2001)

\bibitem{titantah-param}
J.T. Titantah, M.~Karttunen, Eur. Phys. J. B \textbf{86}, 288 (2013)

\bibitem{nose-molphys84}
S.~Nos{\'e}, Mol. Phys. \textbf{52}, 255 (1984)

\bibitem{hoover85}
W.G. Hoover, Phys. Rev. A \textbf{31}, 1695 (1985)

\bibitem{heyi10}
Y.~He, T.~Zeng, J. Phys. Chem. C \textbf{114}, 18023 (2010)

\bibitem{perdew85}
J.P. Perdew, Int. J. Quantum Chem. \textbf{19}, 497 (1985)

\bibitem{HybertsenLouie1986}
M.S. Hybertsen, S.G. Louie, Phys. Rev. B \textbf{34}, 5390 (1986)

\bibitem{hedin65}
L.~Hedin, Phys. Rev. \textbf{139}, A796 (1965)

\bibitem{simard91}
B.~Simard, P.~Hackett, A.~James, P.~Langridgesmith, Chem. Phys. Lett.
  \textbf{186}, 415 (1991)

\bibitem{martin87}
R.L. Martin, J. Chem. Phys. \textbf{86}, 5027 (1987)

\bibitem{Idrobo07}
J.C. Idrobo, W.~Walkosz, S.F. Yip, S.~Oeguet, J.~Wang, J.~Jellinek, Phys. Rev.
  B \textbf{76}, 205422 (2007)

\bibitem{pereiro07}
M.~Pereiro, D.~Baldomir, Phys. Rev. A \textbf{75}, 033202 (2007)

\bibitem{fedrigo93}
S.~Fedrigo, W.~Harbich, J.~Buttet, J. Chem. Phys. \textbf{99}, 5712 (1993)

\bibitem{hakkinen04}
H.~H{\"a}kkinen, M.~Moseler, O.~Kostko, N.~Morgner, M.~Hoffmann, B.~von
  Issendorff, Phys. Rev. Lett. \textbf{93} (2004)

\bibitem{kuisma15}
M.~Kuisma, A.~Sakko, T.P. Rossi, A.H. Larsen, J.~Enkovaara, L.~Lehtovaara,
  T.~Rantala, Phys. Rev. B \textbf{91} (2015)

\bibitem{evanoff05}
D.~Evanoff, G.~Chumanov, ChemPhysChem \textbf{6}, 1221 (2005)

\bibitem{kinnan10}
M.K. Kinnan, G.~Chumanov, J. Phys. Chem. C \textbf{114}, 7496 (2010)

\bibitem{lin11}
Q.~Lin, Z.~Sun, Optik \textbf{122}, 1031 (2011)

\bibitem{sendova06}
M.~Sendova, M.~Sendova-Vassileva, J.~Pivin, H.~Hofmeister, K.~Coffey,
  A.~Warren, J. Nanosci. Nanotechnol. \textbf{6}, 748 (2006)

\bibitem{atay04}
T.~Atay, J.~Song, A.~Nurmikko, Nano Lett. \textbf{4}, 1627 (2004)

\bibitem{weissker11}
H.C. Weissker, C.~Mottet, Phys. Rev. B \textbf{84} (2011)

\bibitem{chakraborty14}
I.~Chakraborty, J.~Erusappan, A.~Govindarajan, K.S. Sugi, T.~Udayabhaskararao,
  A.~Ghosh, T.~Pradeep, Nanoscale \textbf{6}, 8024 (2014)

\bibitem{kumar10}
S.~Kumar, M.D. Bolan, T.P. Bigioni, J. Am. Chem. Soc. \textbf{132}, 13141
  (2010)

\bibitem{lunskens15}
T.~Luenskens, P.~Heister, M.~Thaemer, C.A. Walenta, A.~Kartouzian, U.~Heiz,
  Phys. Chem. Chem. Phys. \textbf{17}, 17541 (2015)

\bibitem{halder15}
A.~Halder, R.~Chattopadhyay, S.~Majumder, S.~Bysakh, M.C. Paul, S.~Das, S.K.
  Bhadra, M.~Unnikrishnan, Appl. Phys. Lett. \textbf{106}, 011101 (2015)

\bibitem{zheng07}
J.~Zheng, P.R. Nicovich, R.M. Dickson, Annu. Rev. Phys. Chem. \textbf{58}, 409
  (2007)

\bibitem{link00}
S.~Link, M.~El-Sayed, Int. Rev. Phys. Chem. \textbf{19}, 409 (2000)

\bibitem{wiley07}
B.J. Wiley, Y.~Chen, J.M. McLellan, Y.~Xiong, Z.Y. Li, D.~Ginger, Y.~Xia, Nano
  Lett. \textbf{7}, 1032 (2007)

\end{thebibliography}

\end{document}